\documentclass{article}    
\usepackage{graphicx}
\usepackage{amssymb}

\begin{document}           

\title{Self-image Effects in Diffraction and Dispersion}  
\author{
H.C. Rosu$^1$, J. P. Trevi\~no$^1$, H. Cabrera$^1$, 
and J. S. Murgu\'{\i}a$^2$ \\ 
{$^1$\it  \small IPICyT - Instituto Potosino de Investigaci\'on Cient\'{\i}fica y Tecnol\'ogica,}\\
{\it \small Apdo Postal 3-74 Tangamanga, 78231 San Luis Potos\'{\i}, M\'exico.}\\
{$^2$ \it \small 
Universidad Aut\'onoma de San Luis Potos\'{\i}, 87545 San Luis
Potos\'{\i}, SLP. M\'exico.}\\
 {\it \small emails:  hcr, jpablo, cabrera@ipicyt.edu.mx, ondeleto@uaslp.mx }}

\maketitle                 

\noindent Talbot's self-imaging effect occurrs in near-field
diffraction. In the rational paraxial approximation, the Talbot
images are formed at distances $z=p/q$, where $p$ and $q$ are
coprimes, and are superpositions of $q$ equally spaced images of the
original binary transmission (Ronchi) grating. This interpretation
offers the possibility to express the Talbot effect through Gauss
sums. Here, we pay attention to the Talbot effect in the case of
dispersion in optical fibers presenting our considerations based on
the close relationships of the mathematical representations of
diffraction and dispersion. Although dispersion deals with
continuous functions, such as gaussian and supergaussian pulses,
whereas in diffraction one frequently deals with discontinuous
functions, the mathematical correspondence enables one to
characterize the Talbot effect in the two cases with minor
differences. In addition, we apply the wavelet transform to the
fractal Talbot effect in both diffraction and fiber dispersion. In
the first case, the self similar character of the transverse
paraxial field at irrational multiples of the Talbot distance is
confirmed, whereas in the second case it is shown that the field is
not self similar for supergaussian pulses. Finally, a high-precision
measurement of irrational distances employing the fractal index
determined with the wavelet transform is pointed out.


\section{Introduction}

Near field diffraction can produce images of periodic structures
such as gratings without any other means. This is known since 1836
when this self-imaging phenomenon has been discovered by H.F.
Talbot, one of the inventor pioneers of photography \cite{hft}. Take
for example a periodic object as simple as a Ronchi grating which is
a set of black lines and equal clear spaces on a plate repeating
with period $a$. In monochromatic light of wavelength $\lambda$ one
can reproduce its image at a ``focal" distance known as the Talbot
distance given by $z_{\rm T}=a^2 \lambda ^{-1}$, a formula first
derived by Lord Rayleigh in 1881 \cite{Rayleigh81}. Moreover, more
images show up at integer multiples of the Talbot distance. It was
only in 1989 that the first and only one review on the Talbot effect
has been written by Patorski \cite{pator}.

In the framework of Helmholtz equation approach to the physical
optics of a $\delta$-comb grating, Berry and Klein \cite{ref1}
showed in 1996 that in rational Talbot planes, i.e., planes situated
at rational distances in units of the Talbot distance, the paraxial
diffraction wavefield has a highly interesting arithmetic structure
related to Gauss sums and other fundamental relationships in number
theory. Moreover, they showed that at irrational distances a fractal
structure of the diffraction field can be proved with sufficient
experimental evidence.

Here, after briefly reviewing the results of Berry \& Klein, we show
that analogous results can be obtained easily for the case of
dispersion in linear optical fibers. Moreover, we apply the wavelet
transform \cite{primer} to the fractal Talbot problem. The point
with the wavelet transform is that it contains more information with
respect to the Fourier transform, which is behind the Helmholtz
equation. Wavelet transforms have been considered as a very suitable
mathematical microscope for fractals due to their ability to reveal
the construction rules of fractals and to resolve local scaling
properties as noticed before for the case of fractal aggregates
\cite{arneodo88}.


\section{Rational Talbot effect for Ronchi gratings}

The diffraction due to Ronchi gratings can be approached
analytically using the Helmholtz equation. Passing to dimensionless
transverse and paraxial variables $\xi=x/a$ and $\zeta = z/a$,
respectively, the scalar wave solution $\Psi (\xi , \zeta)$ of the
Helmholtz equation can be expressed as a convolution in $\xi$ of the
Ronchi unit cell square function
\begin{equation}\label{unit cell}
g(\xi) =\left\{\begin{array}{c}
1\quad \xi\in [-\frac{1}{4},\frac{1}{4}]\\
$\, \,$0\quad \xi \ni [-\frac{1}{4},\frac{1}{4}]~,
\end{array}\right.
\end{equation}
and the Dirac comb transmittance, i.e.
\begin{equation}\label{scalarH}
\Psi(\xi , \zeta) =\int _{-1/2}^{+1/2} g(\xi ^{'})\left(\sum
_{n=-\infty}^{\infty} \exp [i \, 2\pi n(\xi -\xi ^{'})]\exp [i\Theta
_n(\zeta)]\right)d\xi ^{'}~.
\end{equation}
In the previous formulas, the unit cell is the single spatial period
of the grating, which we take centered at the origin and of length
equal to unity and $\Theta _n(\zeta)=2\pi\zeta \frac{a^2}{\lambda
^2} \sqrt{1-\left(\frac{n\lambda}{a}\right)^2}$ is a phase produced
by the diffraction of the Dirac comb `diagonal' rays. The so-called
Fresnel approximation for this phase is a Taylor expansion up to the
second order for the square root leading to $\Theta _s(\zeta)
\approx -\pi n^2\zeta$. It can be easily shown now that in the
Fresnel approximation Eq.~\ref{scalarH} can be written as an
infinite sum of phase exponentials in both variables $\xi$ and
$\zeta$
\begin{equation}\label{sum_exps}
\Psi_{\rm p}(\xi , \zeta) = \sum _{n=-\infty}^{\infty} g_n \exp [i\,
2\pi n \xi - i\pi n^2\zeta]=\sum _{n=-\infty}^{\infty} g_n\psi _{\rm
p} (\xi ,
\zeta)~, 
\end{equation}
where the amplitudes $g_n$ are the Fourier modes of the
transmittance function of the Ronchi grating
\begin{equation}\label{Fourier_g}
g_n=\int _{-1/4}^{+1/4}\exp [-i \, 2\pi n\xi ^{'}]d\xi ^{'}~.
\end{equation}
Furthermore, by discretizing the optical propagation axis $\zeta$ by
means of rational numbers, one can write the rational paraxial field
as a shifted delta comb affected by phase factors, which is the main
result of Berry and Klein:
\begin{equation}\label{rational}
\psi _{\rm p}\left(\xi , \frac{p}{q}\right)=\frac{1}{q^{1/2}}\sum
_{n=-\infty}^{\infty}\Phi_{\rm diffr}(n;q,p)\,\delta\left(\xi
_{p}-\frac{n}{q}\right), 
\end{equation}
where $\xi _p=\xi -e_p/2$ and $e_p = 0$(1) if $p$ is even (odd). The
factors $\Phi_{\rm diffr}(n;q,p)$ are actually phase factors and
will be specified in the next section. They appear to be the
physical quantities directly connected to number theory.
At the same time, 
this rational approximation allows for the following important
physical interpretation of the paraxial self-imaging process: {\em
in each unit cell of the plane $p/q$, $q$ images of the grating
slits are reproduced with spacing $a/q$ and intensity reduced by
$1/q$}.


\section{Rational Talbot effect in linear optical fibers}
As known, in fiber optics technology, electromagnetic dispersion is
defined in terms of the {\em propagation constant} (wavenumber) of
each frequency mode $\beta(\omega) =n(\omega) \frac{\omega}{c}$. In
the following we will use one of the simplest optical fibers having
a core-cladding step profile of the index of refraction. In
addition, the famous slowly varying envelope approximation
(henceforth SVEA) is a realistic approach when treating the
propagation of quasi-monochromatic fields, such as laser fields and
other types of coherent beams within such materials. For more
details we refer the reader to textbooks \cite{agr}.
\medskip


{\em SVEA} means decomposing the electromagnetic fields in two
factors: a rapidly varying phase component and a slowly varying
amplitude field $A$ enveloping the rapid oscillatory fields. The
following Schr\"odinger-like dispersion equation can be obtained for
$A$ in the {\em SVEA} approximation
\begin{equation}\label{eq:prop4}
2i\frac{\partial A}{\partial z}=-{\rm sign}(\beta_2)\frac{\partial^2
A}{\partial {\tilde t}^2}~,
\end{equation}
where $\beta _2$ is the second coefficient in the Taylor expansion
of the propagation constant in the neighbourhood of the central
resonance frequency. This is the simplest form of the dispersion
equation that one can envision in which actually no material
propagation shows up. It can be fulfilled in the practical situation
when the dielectric medium has sharp resonances ($\delta \omega
_r\ll \omega _r$). Because of technological requirements, $\beta _2$
is usually  a negative parameter corresponding to the so-called
anomalous dispersion region. As can be seen, the SVEA equation has
exactly the same mathematical form as the diffraction equation in
the paraxial approximation:
\begin{equation}\label{eq:fresprop}
2i\frac{\partial \Psi _{\rm p}}{\partial z}=\frac{\partial^2 \Psi
_{\rm p}}{\partial x^2}~,
\end{equation}
where $\Psi _{\rm p}$ is the electric field close to the propagation
axis.



Many results in diffraction can be translated to the case of
dispersion in fibers by using the following substitutions
\[
\begin{array}{rcl}
x&\rightarrow&\tilde t \quad (\xi \rightarrow \tau)\\
y&\rightarrow&{\bf r} \\
z&\rightarrow&z \quad (\zeta \rightarrow \zeta).
\end{array}
\]
In the first row one passes from the grating axis to the time axis
of a frame traveling at the group velocity of a pulse. In the second
row one passes from the second grating axis that here we consider
constant to the transverse section of the optical fiber. Finally,
the propagation axis remains the same for the two settings. This
{\it change of variables} will be used here to compare the results
obtained in the two frameworks.

The general solution of the {\em SVEA} dispersion equation
~(\ref{eq:prop4}) for the amplitude $A(z,\tilde t)$ depends on the
initial conditions. Assuming a periodic input signal of period $T$
written as a Fourier series, i.e., $A(0,\tilde t)=\sum
_{n=-\infty}^{n=\infty}C_n^0e^{-i\omega _n \tilde t}$, where $C_n^0$
are the Fourier coefficients of the initial pulse at the entrance of
an optical fiber with linear response, one can write the pulse at an
arbitrary $z$ as follows:
\begin{equation}\label{disp_ampl_solgral}
A(z,\tilde t)=\sum C^0_n\exp{\left[i\frac{\omega^2_nz}{2}-i\omega_n
\tilde t\right]} \qquad \mbox{where  }\omega_n=2\pi n/T.
\end{equation}

If the scaling of variables $\tau=\tilde t/T$, $\zeta=2z/z_{{\rm
T}}$
is employed, $A(z,\tilde{t})$  can be rewritten as
\begin{equation}\label{eq:a}
A(\zeta,\tau)=\sum C^0_n\exp{\left[i\pi n^2\zeta-i2\pi
n\tau\right]},
\end{equation}
because the Talbot distance corresponding to this case is $z_{\rm
T}=T^2/\pi$. Just as in the context of diffraction, the convolution
of the unitary cell with the propagator can be equally done before
or after the paraxial approximation is employed. We notice that
Eq.~\ref{eq:a} can be also written as
\begin{equation}\label{dispersed_amplitude_integral}
A(\zeta,\tau)=\int^{T/2}_{-T/2}A(0, \tau')\alpha (\zeta,\tau'-\tau)d
\tau '
\end{equation}
since $C^0_n$ are nothing but the Fourier coefficients of the input
signal and where
\begin{equation}\label{Gsum_A}
\alpha (\zeta,\tau)=\sum^{\infty}_{n=-\infty} \exp{\left[i\pi
n^2\zeta-i2\pi n\tau\right]}
\end{equation}
can be thought of as the analog of the paraxial propagator
\cite{ref1}. In this expression, the trick is to turn the continuous
propagation axis into the rational number axis and also to perform
the integer modulo division of $n$ with respect to the rational
denominator of the propagation axis, i.e.,
\begin{equation}\label{rational1}
\zeta=\frac{p}{q},\qquad\quad n=lq+s.
\end{equation}
Through this approximation, the sum over $n$ is divided into two
sums: one over negative and positive integers $l$, and the other one
over $s\equiv n\,\mbox{(mod}\,q)$
\begin{equation}\label{double_sum}
\alpha
\left(\frac{p}{q},\tau\right)=\sum^{\infty}_{l=-\infty}\sum^{q-1}_{s=0}
\exp{\left[i\pi (lq+s)^2\frac{p}{q}-i2\pi (lq+s)\tau\right]}.
\end{equation}
This form of  $\alpha(\zeta,\tau)$ is almost exactly the same as
given by Berry \& Klein \cite{ref1} and by Matsutani and \^Onishi
\cite{ref2}. The difference is that the sign of the exponent is
opposite. Following these authors one can express $\alpha$ in terms
of the Poisson formula leading to
\begin{equation}\label{14}
\alpha \left(\frac{p}{q},\tau\right)=
                               \frac{1}{\sqrt{q}}\sum _{n=-\infty}^{\infty} \Bigg[\frac{1}{\sqrt{q}}\sum _{s=0}^{q-1}
                              \exp{\left[i\pi\left(\frac{p}{q}s^2-2s\tau\right)\right]}\Bigg]\delta \left(\tau _p+\frac{n}{q}\right)~,
\end{equation}
where $\tau _p$ is a notation similar to $\xi _p$. 
We can also write Eq.~\ref{14} in the form
$$
\alpha \left(\frac{p}{q},\tau\right)=
                  \frac{1}{\sqrt{q}}\sum _{n=-\infty}^{\infty}
                  \Phi _{\rm disp}(n;q,p)\, \delta \left(\tau _p+\frac{n}{q}\right)~,
$$
which is similar to Eq.~\ref{rational}. The rest of the calculations
are straightforwardly performed though they are lengthy. By
algebraic manipulations the phase factor can be easily obtained and
we reproduce below the two expressions for direct comparison
\begin{equation} \label{phase1}
{\Phi}_{\rm disp}(n;q,p)=\frac{1}{\sqrt{q}}\sum_{s=0}^{q-1}
\exp{\left\{   i\frac{\pi}{q} \left[ps^2-2s(n-\frac{qe_p}{2})
\right] \right\}}
\end{equation}
\begin{equation} \label{Phase2}
 \Phi _{\rm diffr}(n;q,p)=\frac{1}{\sqrt{q}}\sum_{s=0}^{q-1} \exp{\left\{   i\frac{\pi}{q} \left[ 2s(n+\frac{qe_p}{2})-ps^2  \right] \right\}}~.
\end{equation}
Both phases are special types of Gauss sums from the mathematical
standpoint. The difference of signs here appears because of the sign
convention chosen for the Fourier transform. Not surprisingly, the
changes in the mathematical formulation are minimal although the
experimental setup is quite different. The final results are the
following:

$p$ even, $q$ odd:
\begin{eqnarray*}
\Phi_{{\rm disp}}(n;p,q)&=&\left(\frac{p}{q}\right)_J\exp\left(-
i\frac{\pi}{4}\left[\left(q-1\right)+\frac{p}{q}(2n\bar
p_q)^2\right]\right) ,\\
\Phi_{{\rm diffr}}(n;p,q)&=&\left(\frac{p}{q}\right)_J\exp\left(+
i\frac{\pi}{4}\left[\left(q-1\right)+\frac{p}{q}(2n\bar
p_q)^2\right]\right).
\end{eqnarray*}

\medskip

$p$ odd, $q$ odd:
\begin{eqnarray*}
\Phi_{{\rm disp}}(n;p,q)&=&\left(\frac{p}{q}\right)_J\exp\left(-
i\frac{\pi}{4}\left[\left(q-1\right)+2\bar
2_q^3\frac{p}{q}((2n-q)\bar
p_q)^2\right]\right) ,\\
\Phi_{{\rm diffr}}(n;p,q)&=&\left(\frac{p}{q}\right)_J\exp\left(+
i\frac{\pi}{4}\left[\left(q-1\right)+2\bar
2_q^3\frac{p}{q}((2n+q)\bar p_q)^2\right]\right).
\end{eqnarray*}

\medskip

$p$ odd, $q$ even:
\begin{eqnarray*}
\Phi_{{\rm disp}}(n;p,q)&=&\left(\frac{q}{p}\right)_J\exp\left(-
i\frac{\pi}{4}\left[-p+\frac{p}{q}((2n-q)\bar
p_q)^2\right]\right) ,\\
\Phi_{{\rm diffr}}(n;p,q)&=&\left(\frac{q}{p}\right)_J\exp\left(+
i\frac{\pi}{4}\left[-p+\frac{p}{q}((2n+q)\bar p_q)^2\right]\right).
\end{eqnarray*}
In all these formulas, the so-called Jacobi symbols in number theory
\cite{koblitz} lie in front of the exponentials and the bar notation
defines the inverse in a given modulo class, i.e., $p\bar p_q\equiv
1({\rm mod}\,q)$.


If one tries to make computer simulations using the Fourier
transform method, the Gibbs phenomenon is unavoidable for
discontinuous transmittance functions. However, in the case of fiber
dispersion, one class of continuous pulses one could work with are
the supergaussian ones, i.e., functions of the following form
\begin{equation}\label{superG}
A(\zeta =0,\tau) = A_0\exp \bigg[\frac{-\tau ^N}{\sigma _0}\bigg] ~,  
\end{equation}
where $N$ is any even number bigger than two. The bigger the chosen
$N$ the more the supergaussian pulse resembles a square pulse. In
our simulations we used the fixed value $N=12$.


\section{Irrational Talbot effect}

\subsection{Fractal approach}


In the Talbot terminology the self-reconstructed images in the
planes $z=(p/q)z_{\rm T}$ consist of $q$ superposed copies of the
grating as already mentioned, completed with discontinuities.
Although there is a finite number of images at fractional distances,
they still represent an infinitesimal subset of all possible images
that occur at the irrational distances.


\begin{figure}[x]
\centering
      \includegraphics[height=7.8cm]{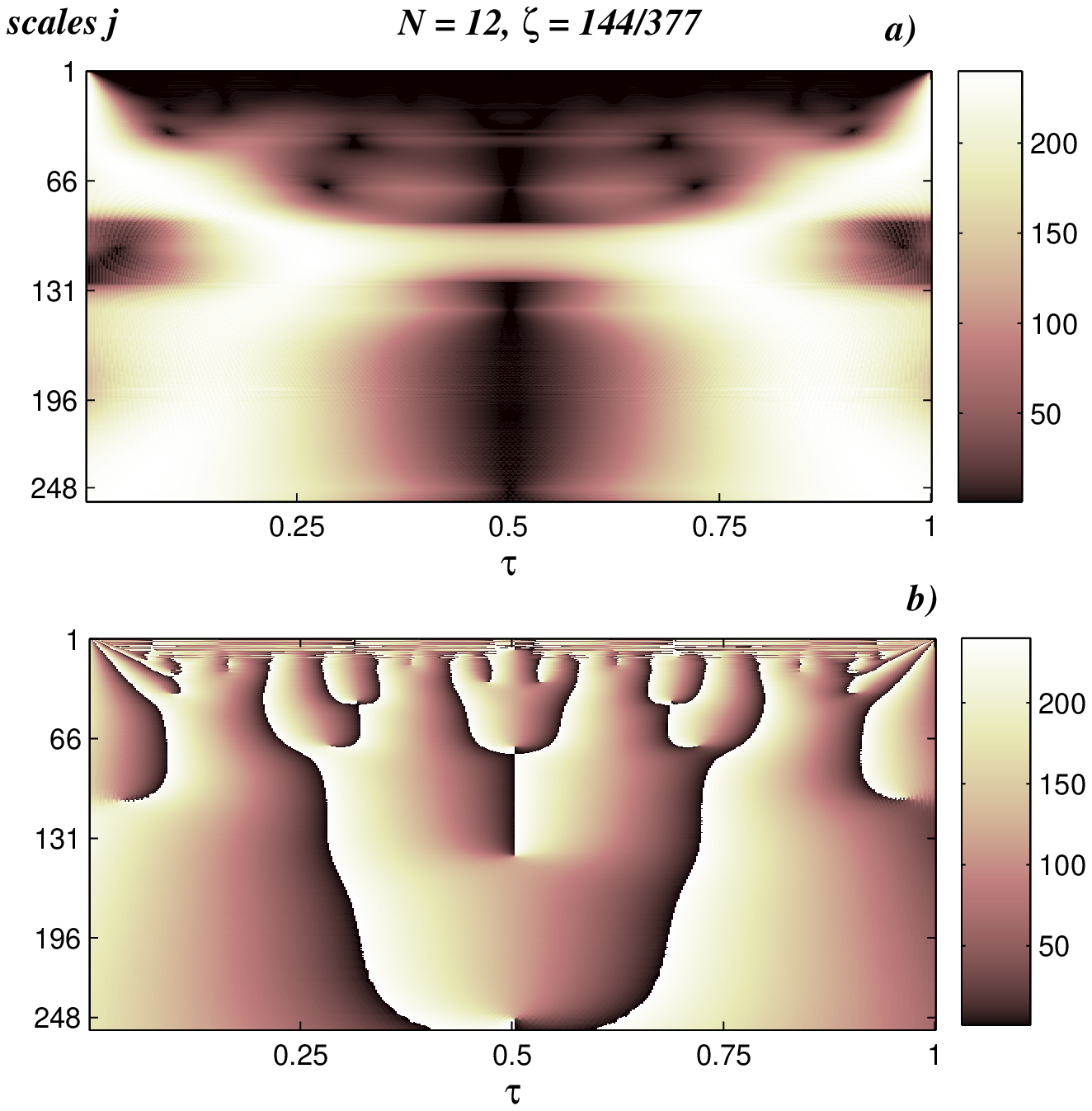}  
  \caption{(Color online) Wavelet representations of the (a) amplitude and (b) phase
of the Talbot dispersed supergaussian field ($N=12$) for $\zeta =
144/377$.}\label{berry_FrTN}
\end{figure}

In the planes located at irrational `fractions' of the Talbot
distance the light intensity is a {\em fractal function} of the
transverse variable. The field intensity has a definite value at
every point, but its derivative has no definite value. Such fractal
functions are described by a fractal dimension, $D$, between one
(corresponding to a smooth curve) and two (corresponding to a curve
so irregular that it occupies a finite area). In the case of  Ronchi
gratings, for example, the fractal dimension of the diffracted
intensity in the irrrational transverse
planes is 3/2 \cite{ref7}. 

To obtain the fractal Talbot images, Berry \& Klein considered the
irrational $\zeta _{\rm irr}$ as the limit $m \rightarrow \infty$ of
a sequence of rationals $p_m/q_m$. In particular, they employed the
successive truncations of the continued fraction for $\zeta _{\rm
irr}$, namely
\begin{equation}\label{bk25}
 \zeta _{\rm irr} =a_0+\frac{1}{a_1+\big[\frac{1}{(a_2+\cdot \cdot \cdot)}\big]}~,
\end{equation}
where the $a_m$ are positive integers. These sequences give best
approximations, in the sense that $p_m/q_m$ is closer to $\zeta
_{\rm irr}$ than any other fraction with denominator $q\leq q_m$. As
a matter of fact, they considered the golden mean $\zeta _{\rm
irr}=(5^{1/2}-1)/2$, for which all $a_m$ are unity, and the $p_m$
and $q_m$ are Fibonacci numbers. Because of the symmetries of the
paraxial field, the Talbot images are the same at $\zeta_{{\rm G}}$
and $\bar{\zeta}_{{\rm G}}\equiv 1-\zeta_{{\rm G}}$, that is
\begin{equation}\label{bk26}
\bar{\zeta}_{{\rm G}}=\frac{3-5^{1/2}}{2}=0.381966... =\lim _{{\rm
\#} \rightarrow \infty} \{ 0~,\, \frac{1}{2}~,\,
\frac{1}{3}~,\,\frac{2}{5}~,\,
\frac{3}{8}~,\frac{5}{13}~,\,\frac{8}{21}~,\, ...\}~.
\end{equation}

Moreover, one can define the so-called {\em carpets} which  are
wave intensity patterns forming fractal surfaces on the $(\xi \, ,
\zeta)$ plane, i.e., the plane of the transverse periodic coordinate
and the propagation
coordinate. 
Since they are surfaces, their fractal dimension takes values
between two and three. According to Berry, in general for a surface
where all directions are equivalent, the fractal dimension of the
surface is one unit greater than the dimension of an inclined curve
that cuts through it. Taking into account that for a Talbot carpet
the transverse image curves have fractal dimension $D=3/2$, then the
carpet's dimension is expected to be 5/2. However, Talbot landscapes
were found not to be isotropic. For fixed $\xi$, as the intensity
varies as a function of distance $\zeta$ from
the grating, the fractal dimension is found to be 7/4, one quarter more than in the transverse case. 
Therefore the longitudinal fractals are slightly more irregular than
the transverse ones. In addition, the intensity is more regular
along the bisectrix canal because of the cancelation of large Fourier components that has fractal dimension of 5/4. 
The landscape is dominated by the largest of these fractal
dimensions (the longitudinal one), and so is a surface of fractal
dimension 1+ 7/4 = 11/4.



\subsection{Wavelet approach}   

Wavelet transforms (WT) are known to have various advantages over
the Fourier transform and in particular they can add up
supplementary information on the fractal features of the signals
\cite{arneodo88}. The one-dimensional WT of $f$ is defined as
follows
\begin{equation}\label{wvl 0a}
W_{f}^{h}({\rm m,n})=\int _{-\infty}^{\infty} f({\rm s}) h^{*}_{{\rm
m,n}}({\rm s})d{\rm s}
\end{equation}
where $h_{{\rm m,n}}$ is a so-called daughter wavelet that is
derived from the mother wavelet $h({\rm s})$ by dilation and shift
operations quantified in terms of the dilation (m) and shift (n)
parameters:
\begin{equation}\label{wvl 0b}
h_{{\rm m,n}}({\rm s})=\frac{1}{\sqrt{{\rm m}}}h\left(\frac{{\rm
s}-{\rm n}}{{\rm m}}\right)
\end{equation}
In the following, we shall use the Morlet wavelets which are derived
from the typical Gaussian-enveloped mother wavelet which is itself a
windowed Fourier transform
\begin{equation}\label{wvl 0c}
h({\rm s})=\exp[-({\rm s}/{\rm s}_0)^2]\exp (i2\pi k{\rm s}).
\end{equation}
The point is that if the mother wavelet contains a harmonic
structure, e.g., in the Morlet case the phase $\exp(i2\pi k{\rm
s})$), the WT represents both frequency and spatial information of
the signal.

In the wavelet framework one can write the expansion of an arbitrary
signal $\varphi(t)$ in an orthonormal wavelet basis in the form
\begin{equation}\label{wvl_1}
\varphi(t)=\sum _{\rm m}\sum _{\rm n} \varphi _{\rm n}^{\rm m} h
_{{\rm m,n}}(t)~,
\end{equation}
i.e., as an expansion in the dilation and translation indices, and
the coefficients of the expansion are given by
\begin{equation}\label{wvl_2}
\varphi _{\rm n}^{\rm m}=\int _{-\infty}^{\infty} \varphi(t)h _{{\rm
m,n}}(t)dt~.
\end{equation}
The orthonormal wavelet basis functions $W _{{\rm m,n}}(t)$ fulfill
the following dilation-translation property
\begin{equation}\label{wvl_3}
h _{{\rm m,n}}(t)=2^{{\rm m}/2}h (2^{\rm m}t-{\rm n})~.
\end{equation}

In the wavelet approach the fractal character of a certain signal
can be inferred from the behavior of its power spectrum $P(\omega)$,
which is the Fourier transform of the autocovariance (also termed
autocorrelation) function and in differential form
$P(\omega)d\omega$ represents the contribution to the variance of a
signal from frequencies between $\omega$ and $\omega+d\omega$.
Indeed, it is known that for self-similar random processes the
spectral behavior of the power spectrum is given by \cite{wor,stas}
\begin{equation}\label{wvl_4}
P_{\varphi}(\omega)\sim |\omega|^{-\gamma _f}~,
\end{equation}
where $\gamma _f$ is the spectral parameter of the wave signal. 
In addition, the variance of the wavelet coefficients possesses the
following behavior \cite{stas}
\begin{equation}
{\rm var} \,\varphi _{\rm n}^{\rm m} \approx \left(2^{{\rm
m}}\right)^{-\gamma _f}~.
\end{equation}

These formulas are certainly suitable for the Talbot transverse
fractals because of the interpretation in terms of the regular
superposition of identical and equally spaced grating images. We
have used these wavelet formulas in our calculations related to the
same rational paraxiallity for the two cases of transverse
diffraction fields (Fig.~1) and the fiber-dispersed optical fields
(Fig.~2), respectively. The basic idea is that the above-mentioned
formulas can be employed as a checking test of the self-similarity
structure of the optical fields. The requirement is to have a
constant spectral parameter $\gamma _f$ over many scales. In the
case of supergaussian pulses, their dispersed fields turned out not
to have the self-similarity property as can be seen by examining
Fig.~2 where one can see that the constant slope is not maintained
over all scales. In Figs.~3 and 4 the behavior of the wavelet
transform using Morlet wavelets for the diffraction field is
displayed. A great deal of details can be seen in all basic
quantities of the diffracted field, namely in the intensity,
modulus, and phase. On the other hand, the same wavelet transform
applied to the N=12 supergaussian dispersed pulse (see Fig.~5),
although showing a certain similarity to the previous discontinuous
case, contains less structure and thus looks more regular. This
points to the fact that if in diffraction experiments one uses
continuous transmittance gratings the fractal behavior would turn
milder.



\section{Conclusion}

The fractal aspects of the paraxial wavefield have been probed here
by means of the wavelet transform for the cases of diffraction and
fiber dispersion. In the case of diffraction, the previous results
of Berry and Klein are confirmed showing that the wavelet approach
can be an equivalent and more informative tool. The same procedure
applied to the case of fiber dispersion affecting the paraxial
evolution of supergaussian pulses indicates that the self-similar
fractal character does not show up in the latter type of axial
propagation. This is a consequence of the continuous transmittance
function of the supergaussian pulses as opposed to the singular one
in the case of Ronchi gratings.

As a promising perspective, we would like to suggest the following
experiment by which irrational distances can be determined. The idea
is that the spectral index of the Talbot fractal images can be used
as a very precise pointer of rational and irrational distances with
respect to the Talbot one. Suppose that behind a Ronchi grating
under plane wave illumination a CCD camera is mounted axially by
means of a precision screw. The Talbot image at $z_{\rm T}$ can be
focused experimentally and can be used to calibrate the whole
system. An implemented real time wavelet computer software can
perform a rapid determination of the fractal index $\gamma _f$,
which in turn allows the detection of changes of the distance in
order to determine if the CCD camera is at rational or irrational
multiples of the Talbot distance. Supplementary information on the
irrational distances may be obtained from the amplitude-phase
wavelet representations. To the best of our knowledge, we are not
aware of another experimental setup in which irrational distances
can be determined in such an accurate way. This also points to
high-precision applications in metrology.

Finally, we mention that a version of this work with more
mathematical details will appear soon \cite{besan}.

\section{Acknowledgements}

The first author wishes to thank Dr. Michel Planat for encouraging
him to study the Talbot effect from the modern perspective of Berry
and Klein. The second author would like to thank Dr. V. Vyshloukh
for introducing him to the research of the self image phenomena.
This work was partially sponsored by grants from the Mexican Agency
{\em Consejo Nacional de Ciencia y Tecnolog\'{\i}a} through project
No. 46980.


\newpage

\begin{center}  {\Large First Four Figure Captions} \end{center}

\bigskip

\noindent Fig. 1: (a) The fractal Talbot light intensity $|\Psi
_{\rm p}|^2$ at the twelfth Fibonacci fraction $\zeta
=144/377\approx 0.381963$, which is already `very close' to
$\bar{\zeta} _{\rm G}$ and (b) the plot of the logarithmic variance
of its wavelet coefficients (Eq.~\ref{wvl_2}). The line of negative
slope of the latter semilog plot indicates fractal behavior of the
diffraction wavefield as we expected. The fractal coefficient is
given by the slope and its calculated value is $\gamma _f$.

\medskip

\noindent Fig. 2: Snapshot of the dispersed supergaussian pulse  for
$N=12$ at $\zeta =144/377$.
The log variance plot is monotonically decreasing 
displaying a plateau indicating a nonfractal behaviour of the $N=12$
supergaussian pulse train.

\medskip

\noindent Fig. 3: The wavelet transform of the intensity $|\Psi
_{{\rm p}}|^2$ at fixed $\zeta =144/377$ for (a) the unit cell and
(b) half-period displaced grating unit cell. There is no difference
because the square modulus is plotted.

\medskip

\noindent Fig. 4: Wavelet representations of: (a) the squared
modulus of the amplitude and (b) phase of the Talbot diffraction
field for fixed $\zeta = 144/377$ and a displaced unit cell.

\medskip


\end{document}